Quantized Phonon Spectrum of Single-Wall Carbon Nanotubes

J. Hone[1*], B. Batlogg[2], Z. Benes[3], A.T. Johnson[1] and J.E. Fischer[3]

1. Department of Physics and Astronomy and Laboratory for Research on the Structure of Matter, University of Pennsylvania, Philadelphia PA 19104-6272

2. Bell Laboratories, Lucent Technologies, Murray Hill, NJ 07974

3. Department of Materials Science and Engineering and Laboratory for Research on the Structure of Matter, University of Pennsylvania, Philadelphia PA 19104-6272

Abstract:

The electronic spectra of carbon nanotubes and other nanoscale systems are quantized due to their small radii. Similar quantization in the phonon spectra has been difficult to observe due to the far smaller energy scale. We probed this regime by measuring the temperature-dependent specific heat of purified single-wall nanotubes. The data show direct evidence of 1D quantized phonon subbands. Above 4 K, they are in excellent agreement with model calculations of individual nanotubes, and differ markedly from the specific heat of two-dimensional graphene or three-dimensional graphite. Detailed modeling yields an energy of 4.3 meV for the lowest quantized phonon subband, and a tube-tube (or ÒlatticeÓ) Debye energy of 1.1 meV, implying a small inter-tube coupling in bundles.

* Present address: Department of Physics, California Institute of Technology
† Author to whom correspondence should be addressed

The electronic structure of single-wall carbon nanotubes (SWNTs) has been extensively studied, and is known to reflect confinement of electron waves to the one-dimensional molecular cylinder. However, the low energy phonon structure of SWNTs is largely unexplored experimentally despite considerable theoretical work. The low-energy phonons are related to the mechanical properties and define the thermal conductivity (*1, 2*), which will determine whether applications such as thermal management in molecular electronics are feasible. In addition, detailed knowledge of the phonon structure is important for understanding electron-phonon scattering in nanotubes (*3, 4*).

The phonon spectrum in SWNTs should display quantum size effects, whereby the two-dimensional phonon bands of graphene fold into a set of quantized one-dimensional subbands, as is seen in the electronic band structure (*5, 6*). In reduced dimensions, the fundamental physics of phonon scattering (*7, 8*) and thermal equilibration is changed dramatically. These basic issues, as well as potential applications such as highly sensitive bolometry, have driven continuing interest in low-dimensional phonon systems (*8*). Most experimental work in this field, such as the recent measurement of the quantum of thermal conductance (*9*), has employed artificially designed nanostructures that are suspended to reduce thermal coupling to the substrate. In contrast to such structures, in which the one-dimensional phonon regime has been difficult to access because of the small energy scale (*10*), carbon nanotubes are a molecular system whose small size and high stiffness result in a much larger energy splitting between 1D phonon subbands.

In SWNTs the specific heat at constant pressure $C_P$ is a direct probe of the phonon energy spectrum, the electronic contribution being negligible (*11*). In an ordinary 3D solid, the low-temperature phonon $C_P(T)$ increases as $T^3$. In an isolated nanotube, all of the circumferential degrees of freedom are frozen out at low temperature, so that the phonons are strictly one-dimensional and $C_P(T)$ is linear in T (*11, 12*). However, in a bulk sample, strong phonon coupling between neighboring tubes will lead to 3D behavior and obscure the signature of 1D confinement. Because the phonon contribution to $C_P$ is determined by the phonon (vibrational) density of states (PDOS) as a function of energy, we first examine the 1D PDOS of a SWNT and the effect of intertube coupling (Fig. 1) to set the stage for discussing our results.

The PDOS spectrum is shown for a 1.25 nm diameter nanotube (the average diameter in available material) based on the phonon dispersion calculated by Saito et al. (*13*) (Fig. 1A). An isolated nanotube has a 1D phonon structure with four acoustic branches (one longitudinal, two transverse and one torsional) with linear dispersions $E = \hbar v q$ ($E$ is the phonon energy, $v$ the phonon velocity, and $q$ the wave-vector) (*14*). The periodic boundary condition on the circumferential wave-vector splits each of these modes into 1D ÒsubbandsÓ which translate into the sharp spikes, or 1D van Hove singularities, in the PDOS. The approximate location of the first optical ($\omega > 0$ at $q = 0$) subband is given by $E_{sub} \approx \hbar v/R$ (*11*), where R is the radius of the nanotube. It is clear why SWNTs are ideal for studying low-dimensional phonons: a small R and large $v$ (of order $10^4$ m/s) lead to a measurably large subband splitting (larger nanotubes will have a smaller subband splitting and approach strictly 2D behavior as R increases). Detailed calculations for a 1.25 nm diameter tube predict that the first subband edge is at $E_{sub}$ = 2.7 meV, or 30 K.

In contrast to the 1D PDOS of the nanotube, the calculated PDOS of a single 2D graphene sheet (*15*) (Fig. 1A) varies smoothly, with no 1D singularities. It is greater in magnitude at E = 0 than that of the isolated tube, because a graphene sheet is weak to bending, whereas a tube is markedly stiffer. The acoustic Òlayer bendingÓ mode in graphene has quadratic dispersion $E \propto q^2$ rather than the linear dispersion typical of acoustic modes in 3D solids. The quadratic dispersion yields a constant PDOS in 2D, which dominates the contribution of the other two (linear-dispersing) acoustic modes.

Because SWNTs are found in bundles (ÒropesÓ) of tens to hundreds of tubes, one must consider the effect of inter-tube coupling on the phonon structure and $C_P$. We first examine the analogous situation when graphene sheets are stacked to make 3D graphite. Coupling between adjacent graphene sheets introduces phonon dispersion in the c-direction, shifting spectral weight from lower- to higher-energy states (Fig. 1). The characteristic energy for this process is the c-axis Debye energy near 12 meV. $C_P$ of graphite shows a broad transition from 3D behavior below 12 meV (roughly 150 K) to 2D behavior above. A similar dimensional crossover should occur in SWNT ropes due to tube-tube coupling: the phonon structure of a rope will be 3D at low energy, and reflect the structure of constituent tubes at higher energies. The characteristic energy of this crossover is the transverse Debye energy $E_D^{\perp}$.

A conceptual phase diagram (Fig. 1B) shows how the relationship of $E_D^{\perp}$ to $E_{sub}$ will determine the conditions for dimensional crossover in a nanotube rope. In an isolated tube ($E_D^{\perp}$ = 0), the phonons are 1D at the lowest temperatures: only the four acoustic subbands are occupied. At a temperature $T_{1D}$ ¯ $E_{sub}/6k_B$ (5K), the first optical subband begins to contribute to the specific heat; well above $T_{1D}$, many subbands are occupied and the tube is essentially 2D. In a weakly-coupled rope ($E_D^{\perp} < E_{sub}$), as the temperature increases from zero, the rope will go from a 3D coupled-tube regime to a 1D regime before crossing over to a 2D (multi-subband) regime above $T_{1D}$. If, however, the tubes are strongly coupled ($E_D^{\perp} > E_{sub}$), then the transition will be from 3D directly to 2D, with no evidence of 1D quantization. Using a model based on the compression and shear behavior of graphite, Mizel et al. (*16*) calculated $E_D^{\perp}$ ~ 5 meV (60 K), implying that SWNT ropes are in the strong-coupling regime.

We now turn to the experiment (*17*). SWNT samples were obtained from purified SWNT suspensions. Structural and chemical analysis confirmed that the average tube diameter was 1.25 nm, that the tubes were found in crystalline bundles, and that a small amount (2 at. %) of Ni/Co catalyst remained. The samples were protected from atmospheric contamination. Heat capacity was measured from 300 K to 2 K using a relaxation technique. Two samples (9.5 mg and 2.5 mg) were measured with similar results, implying that there were no significant systematic offsets in the measurement. The smaller mass causes the 2.5 mg data to have higher uncertainty, so we focus on the data from the 9.5 mg sample. The data presented below (Figs. 2, 3, 4) have not been smoothed, so the spread in data points at a given temperature reflects the measurement uncertainty, which is never more than a few percent.

The measured $C_P$, taken on slow cooling from 300 K to 2 K, decreases monotonically with decreasing T, the lowest data point being 0.3mJ/g-K at 2 K (Figs. 2A and 2B, run 1). After run 1 the sample was left overnight at 4.2 K and then measured on heating (Fig. 2B). The heat capacity from 2 to 20 K had increased dramatically, but this ÒexcessÓ heat capacity disappeared above 20 K, consistent with adsorption of helium that had diffused into the vacuum space overnight. High-surface area carbons (including nanotubes (*18*)) are known to adsorb helium, and recent theoretical work (*19*) has predicted a high specific heat for helium adsorbed into the interstitial channels of a SWNT rope. As a check, the sample was warmed to 77 K, pumped out overnight, and then cooled quickly. The fast-cooling data (run 2 in Fig. 2B) are identical to the slow-cooling data of run 1, and reflect the intrinsic specific heat of the sample.

The $C_P(T)$ curves calculated from the theoretical PDOS spectra are shown in Fig. 3. $C_P(T)$ directly reflects the dimensionality: at low temperature, an acoustic phonon mode in d dimensions with dispersion $E(q) \propto q^{\alpha}$ has $C_P \propto T^{d/\alpha}$. $C_P(T)$ of 2D graphene is dominated by the quadratic layer-bending mode, and therefore has a roughly linear T-dependence. In contrast, $C_P$ for 3D graphite decreases more rapidly as T decreases below 80 K, a consequence of the 2D to 3D dimensionality crossover driven by interlayer coupling. Measurements on graphite (*20*) agree with the calculated phonon $C_P$ down to 5K, below which a small electronic contribution causes the measured data to lie slightly above the phonon curve. An isolated nanotube, with linear acoustic bands in 1D, will have $C_P \propto T$ at low T, with an increase in slope due to the contribution from the first subband above $T_{1D} \sim 5$ K. The nanotube curve lies well below the graphene one because the tube has no low-energy counterpart to the layer bending modes. The $C_P$ of a nanotube rope should follow the single-tube curve at high T, then show dimensional crossover to a stronger T-dependence as T decreases. Compared to the analogous behavior in graphite ($E_D^c \sim 12$ meV), strongly coupled ropes ($E_D^{\perp} = 5$ meV) should begin to deviate below the single-tube curve at $\sim 30$ K. The calculated low-T $C_P$ of a strongly coupled rope (*16*) (Fig. 3) shows a 3D behavior similar to graphite.

The measured specific heat (Fig. 3) is clearly largely consistent with the single-tube model, even though the sample consists mostly of large bundles. At intermediate temperatures (20-100 K), the data lie just above the single-tube prediction. We attribute this small discrepancy to 2 at. % residual catalyst (*21*) (Fig. 3). Adding the catalyst contribution to the single tube model fits the data quite well above 4 K. Below 4K, the data lie significantly below the model curve, which we attribute to the crossover to 3D behavior on cooling. A crossover temperature near 4K is much lower than predicted for strongly coupled tubes. Therefore we conclude that the tubes are only weakly coupled, so that 1D quantum effects are observable.

Figure 5 emphasizes the low temperature regime of 1D phonon confinement. The measured $C_P$ increases linearly with T from 2 to 8 K, at which point the slope increases. This behavior is direct evidence for quantized 1D phonon subbands in nanotubes. However $C_P$ does not extrapolate linearly to zero at T = 0, as expected for isolated tubes. We know the sample contains ropes, and we have evidence that intertube coupling is weak. An improved $C_P(T)$ model, accounting for both the quantized phonon subband structure of individual tubes and weak tube-tube coupling, can be derived from a simplified bundle phonon band structure (Fig. 4, inset). The four acoustic bands are combined into a single fourfold-degenerate band with longitudinal Debye energy $E_D^{\parallel}$ and transverse Debye energy $E_D^{\perp}$. A doubly degenerate optical subband enters at $E_{sub}$ with dispersion $E^2 = (\hbar v q_{\parallel})^2 + (E_{sub})^2$. Because $E_{sub} > E_D^{\perp}$, transverse dispersion of the subband can be ignored. The contribution from the acoustic band, with $E_D^{\parallel}$ = 92 meV (1070 K) and $E_D^{\perp}$ = 1.2 meV (14 K), displays roughly cubic temperature dependence below ~ 2 K, above which the inter-tube modes saturate, and $C_P$ displays the linear behavior characteristic of 1D phonons. The contribution from the first subband, with $E_{sub}$ = 4.2 meV (50 K), is only significant above 8 K. The total of the two contributions fits the data extremely well; deviation of ~ 10% in any of the fitting parameters resulted in a noticeably worse fit.

The experimental on-tube parameters derived from the fit can be compared to theory (*13*). The theoretical acoustic mode velocities translate into an effective Debye energy of 103 meV, slightly higher than our fitted 92 meV. Our fitted $E_{sub}$ is larger than the theoretical single-tube value of 2.7 meV. These discrepancies may arise from inter-tube interactions whereby weak coupling modifies the elastic properties of the constituent tubes. For example, the first phonon subband (the low-energy mode with $E_{2g}$ symmetry at $q = 0$) corresponds to tube flattening; this requires significantly more energy in a rope since tubes are constrained by their neighbors (*22*). The experimental tube-tube coupling, measured by $E_D^{\perp}$ = 1.2 meV, is significantly smaller than the theoretical value of 5 meV (*16*). As a possible explanation, we note that Mizel et al. base their model on coupling constants derived from graphite. However, the planes in graphite are identical, and the lattices of neighboring planes are commensurate. In contrast, neighboring tubes in a rope are most likely not identical. They will in general have different chiral angles (*23*) and diameters (*24*), so that the lattice structure on neighboring tubes will not be commensurate. This in turn implies a dramatic weakening of the corrugation in the inter-tube potential; tubes in a real rope may slide or twist more freely than expected from idealized models.

This observation of quantum size effects on the nanotube phonon spectrum and measurement of the above parameters have implications for applications and the theoretical understanding of nanotubes. The existence of quantized subbands in nanotubes indicates that theoretical and experimental work on low-dimensional phonons in artificial structures (*25*) is applicable to this technologically important material. The measured high on-tube Debye energy confirms, in a bulk sample, the high Young's modulus previously observed for individual tubes (*26*). The weak tube-tube coupling, however, implies that the mechanical strength of SWNT ropes will be relatively poor. It may be necessary to cross-link tubes within a rope, or to separate them completely, in order to realize their near-ideal properties in high-strength composites. On the other hand, weak coupling may be an advantage for high thermal conductivity. Berber, et al. (*2*) find that strong tube-tube coupling decreases the high-temperature thermal conductivity of SWNT bundles by an order of magnitude relative to isolated tubes; weak coupling may imply no significant reduction in the thermal conductivity when tubes are bundled into ropes. Similarly, in composites, the inner tubes in a rope should be relatively unperturbed by the surrounding matrix, which could also be an advantage for high thermal conductivity. The issues of commensurability that were raised as an explanation for the weak tube-tube mechanical coupling also suggest that there will be weak electronic coupling between neighboring SWNTs in a rope (*25*). Finally, our value for the first subband energy, large compared to single-tube force-constant theories, provides information about the effect of inter-tube interactions on single-tube deformation energies. An understanding of this effect is also critically important to obtain correct theoretical values for energy of the radial breathing mode, commonly measured by Raman scattering to determine tube diameters and diameter distributions (*22, 27, 28*), and will have implications for the electronic overlap between neighboring tubes.

Supported by NSF grant DMR-9802560 (JH, ATJ), DOE grant DEFG02-98ER45701 (ZB, JEF), and NSF MRSEC grant DMR-9632598. We thank P. Papanek, D. E. Luzzi, and N. M. Nemes for the $C_P$ calculation of graphite, HRTEM, and EDX analyses respectively; and J.-P. Issi for bringing Ref. 19 to our attention. The SWNT material was graciously provided by D. Colbert (Rice University). We acknowledge helpful conversations with E. J. Mele, M.S. Dresselhaus, and A. Mizel.

FIGURE CAPTIONS

Fig. 1. (A) Theoretical phonon density of states (normalized per carbon atom) for 2D graphene (red, Ref. (*13*)), 3D graphite (green, Ref. (*15*)) and an isolated 1.25 nm diameter SWNT (blue, Ref. (*13*)). Interlayer coupling in graphite shifts spectral weight from lower to higher energies. (B) Conceptual phonon characteristic phase diagram for a SWNT rope. With increasing temperature, isolated tube phonons (zero coupling) cross over from a 1D regime where only acoustic subbands are occupied, to a 2D regime as higher (optic) subbands are populated. This occurs at $T_{1D}$, which goes roughly as the inverse tube radius. In contrast, a bundle of weakly coupled tubes follows the lower dashed line: the phonons are 3D at low temperature, crossing over first to a 1D regime at a temperature which depends on the strength of inter-tube coupling, characterized by the transverse Debye energy $E_D^{\perp}$. If the coupling is strong (upper dashed line), the 1D regime is bypassed, and a quantized phonon spectrum is not observed in $C_P(T)$.

Fig. 2. (A) Specific heat of a sample consisting mainly of SWNT ropes, measured on first cooling from 300 to 2 K (run 1). (B) Low-temperature expansion of (A) (solid dots) and subsequent runs. Open triangles represent a heating run after leaving the sample at 4 K overnight, showing the effects of helium adsorption at 4 K and desorption at 20 K. Open circles (run 2) were recorded during rapid cooling after first warming to 77 K to completely desorb helium; these overlap perfectly with run 1, from which we conclude that helium adsorption is only an issue if the sample is held at 4 K for a long time.

Fig. 3. Log-log plot of data (solid dots) compared with calculations for 2D graphene (solid blue), 3D graphite (dashed blue), isolated tubes (solid green) and strongly coupled ropes (dashed green). The data agree with the isolated tube model down to 5 K, indicating that tube-tube coupling is relatively weak. The agreement is improved at high T (solid red curve) by including the contribution of 2 at. % nickel impurities (black curve). Below 5 K the data fall significantly below the isolated tube prediction.

Fig. 4. Data on an expanded (linear) scale (solid dots), and a fit to an anisotropic two-band Debye model which accounts for weak coupling between SWNTs in a rope (black curve). The contribution from acoustic modes with large on-tube Debye energy $E_D^{\parallel}$ and small transverse Debye energy $E_D^{\perp}$ gives the blue curve which fits the data at low temperatures but lies below the data above 8 K. Including the first 1D subband, approximated as a dispersionless optic branch at $E_{sub}$, adds a contribution given by the red curve. These are combined in the black curve, which fits the data over the entire range. Fitting parameters are given in the text; they imply that in real ropes the coupling and first subband threshold energies are respectively smaller and larger than previously believed.

transmission electron microscopy (HRTEM), and energy-dispersive X-ray spectroscopy (EDX). The XRD showed peaks consistent with a crystalline hexagonal lattice; the position of the (1,0) peak gave an average tube diameter of 1.25 nm, consistent with TEM and XRD analysis of similar samples (*25*). HRTEM confirmed that the sample was composed of the usual large ropes of SWNTs. Finally, EDX revealed that the residual catalyst concentration was ≈ 2 at.%. After analysis, the samples were baked at 300 °C under dynamic vacuum for three days to remove atmospheric contaminants and kept under vacuum until minutes before the measurements were begun.

Before the sample was loaded, a dab of Apiezon N grease was affixed to the cold stage, and the heat capacity of the addenda (stage plus grease) was measured over the entire temperature range (2-300 K). The sample was then removed from vacuum under an inert nitrogen atmosphere, weighed, placed onto the greased stage, and then loaded into the cryostat and evacuated. The total heat capacity of the stage and sample was then measured, and the previously measured heat capacity of the addenda subtracted from the total heat capacity to give the data in Fig. 2.

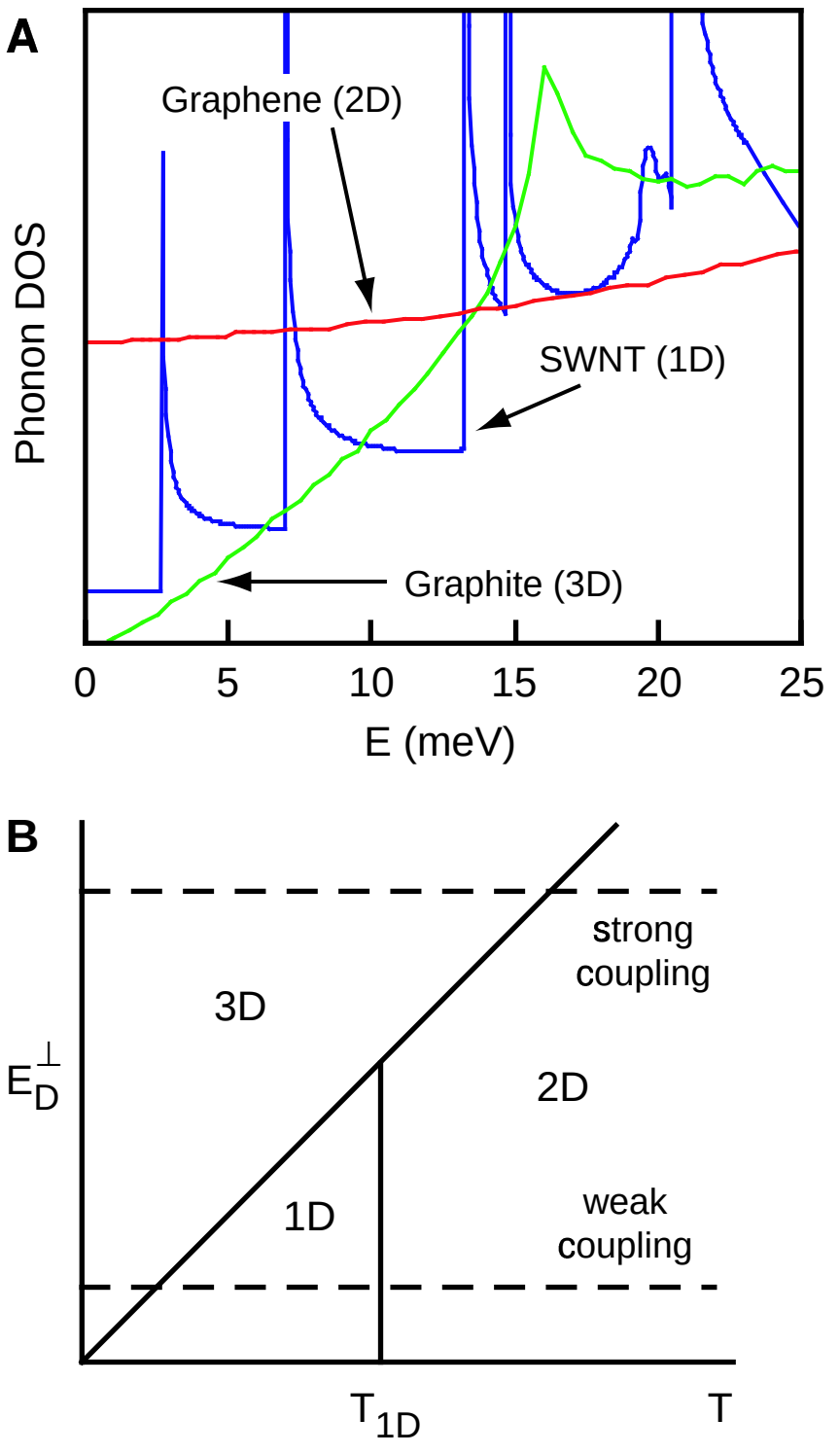

Hone et al.
Fig. 1

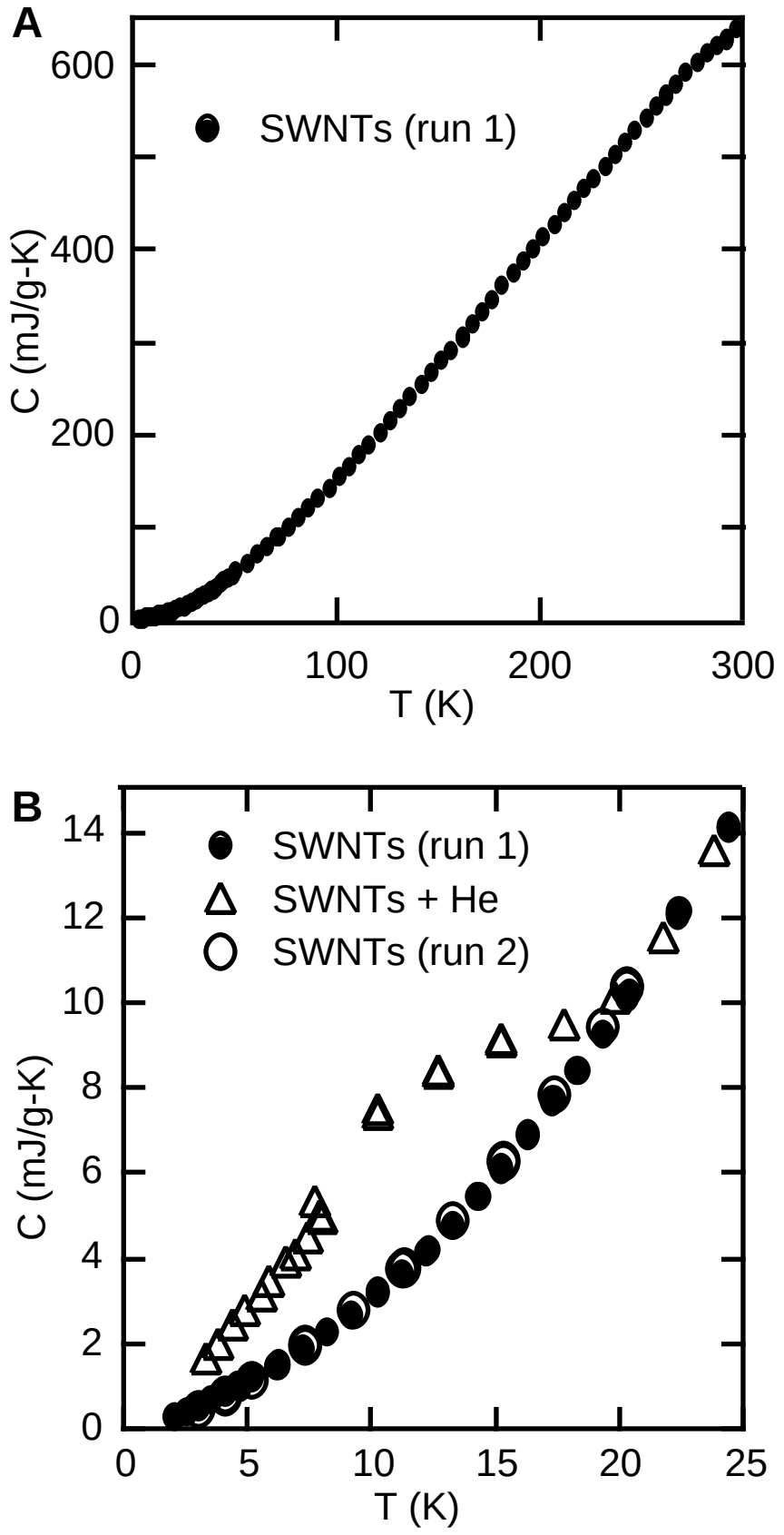

Hone et al.
Fig. 2

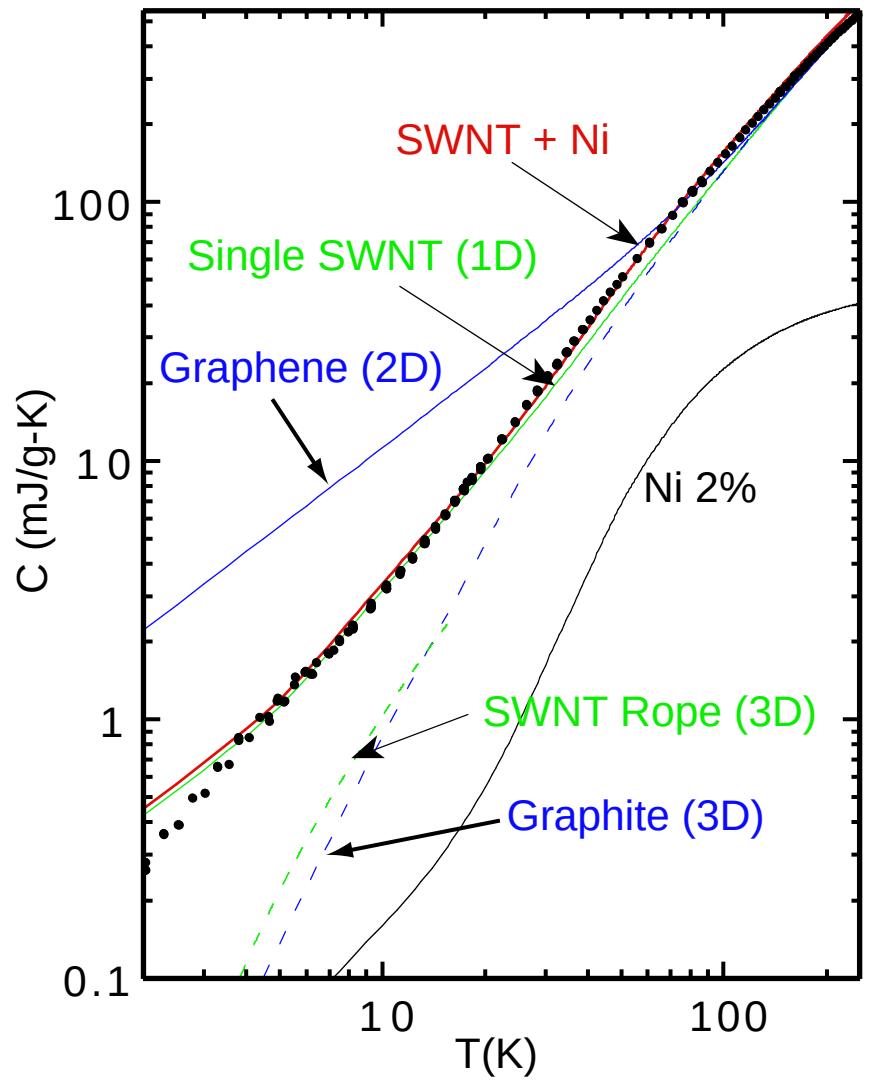

Hone et al.
Fig. 3

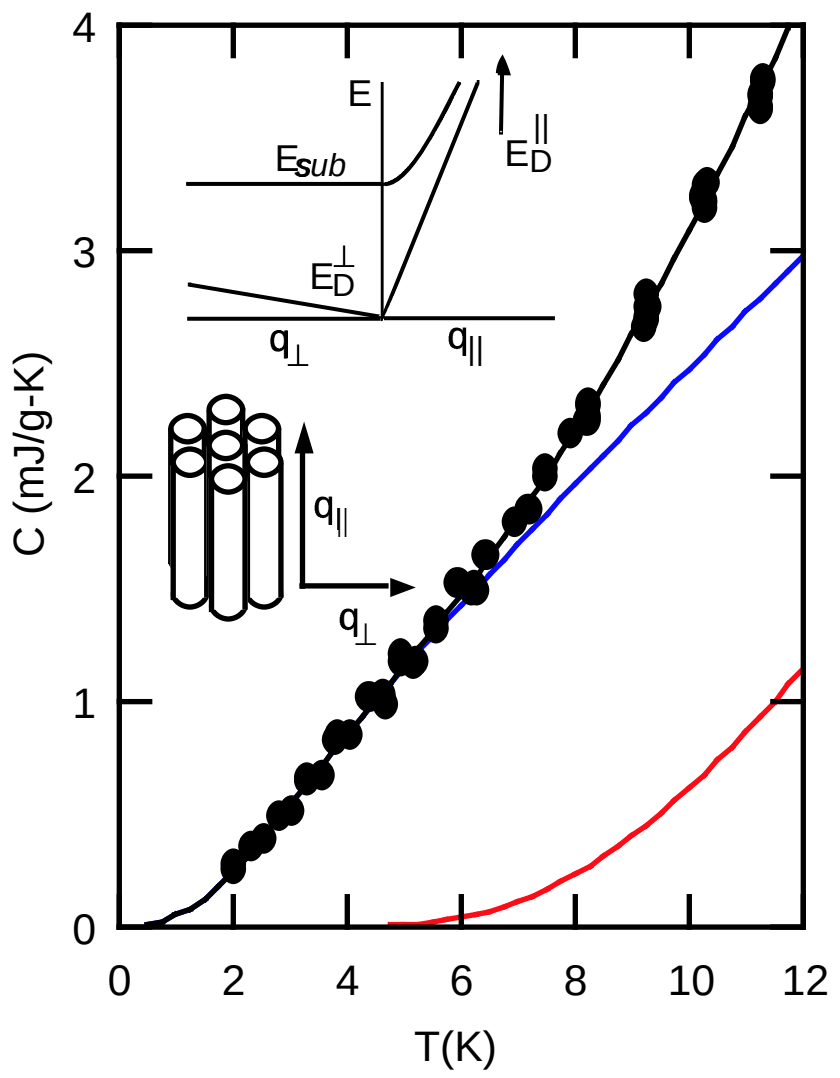

Hone et al.
Fig. 4